\documentclass[12pt]{article}
\usepackage{graphicx}
\def\bra#1{\left\langle #1\right|}
\def\ket#1{\left| #1\right\rangle}
\newcommand{\bers}{\begin{eqnarray*}}
\newcommand{\eers}{\end{eqnarray*}}
\newcommand{\bt}{\begin{itemize}}
\newcommand{\et}{\end{itemize}}
\def\beq{\begin{equation}}
\def\eeq{\end{equation}}
\def\bea{\begin{eqnarray}}
\def\eea{\end{eqnarray}}
\def\nn{\nonumber}
\def\sla#1{\raise.15ex\hbox{$/$}\kern-.57em #1}

\def\sss{\scriptscriptstyle}

\def\barp{{\raise.35ex\hbox
{${\sss (}$}}---{\raise.35ex\hbox{${\sss )}$}}}
\def\bdbarp{\hbox{$B_d$\kern-1.4em\raise1.4ex\hbox{\barp}}}
\def\bsbarp{\hbox{$B_s$\kern-1.4em\raise1.4ex\hbox{\barp}}}

\def\roughly#1{\mathrel{\raise.3ex\hbox
{$#1$\kern-.75em\lower1ex\hbox{$\sim$}}}}

\def\npb#1#2#3{{\it Nucl.\ Phys.} {\bf B#1} (#2) #3}
\def\plb#1#2#3{{\it Phys.\ Lett.} {\bf #1B} (#2) #3}

\def\zpc#1#2#3{{\it Zeit.\ Phys.} {\bf C#1} (#2) #3}

\newread\epsffilein 
\newif\ifepsffileok 
\newif\ifepsfbbfound 
\newif\ifepsfverbose 
\newdimen\epsfxsize 
\newdimen\epsfysize 
\newdimen\epsftsize 
\newdimen\epsfrsize 
\newdimen\epsftmp 
\newdimen\pspoints 
\def\epsfbox#1{\global\def\epsfllx{72}\global\def\epsflly{72}%
 \global\def\epsfurx{540}\global\def\epsfury{720}%
 \def\lbracket{[}\def\testit{#1}\ifx\testit\lbracket
 \let\next=\epsfgetlitbb\else\let\next=\epsfnormal\fi\next{#1}}%
\def\epsfgetlitbb#1#2 #3 #4 #5]#6{\epsfgrab #2 #3 #4 #5 .\\%
 \epsfsetgraph{#6}}%
\def\epsfnormal#1{\epsfgetbb{#1}\epsfsetgraph{#1}}%
\def\epsfgetbb#1{%
\openin\epsffilein=#1
\ifeof\epsffilein\errmessage{I couldn't open #1, will ignore it}\else
 {\epsffileoktrue \chardef\other=12
 \def\do##1{\catcode`##1=\other}\dospecials \catcode`\ =10
 \loop
 \read\epsffilein to \epsffileline
 \ifeof\epsffilein\epsffileokfalse\else
 \expandafter\epsfaux\epsffileline:. \\%
 \fi
 \ifepsffileok\repeat
 \ifepsfbbfound\else
 \ifepsfverbose\message{No bounding box comment in #1; using defaults}\fi\fi
 }\closein\epsffilein\fi}%
\def\epsfclipstring{}
\def\epsfsetgraph#1{%
 \epsfrsize=\epsfury\pspoints
 \advance\epsfrsize by-\epsflly\pspoints
 \epsftsize=\epsfurx\pspoints
 \advance\epsftsize by-\epsfllx\pspoints
 \epsfxsize\epsfsize\epsftsize\epsfrsize
 \ifnum\epsfxsize=0 \ifnum\epsfysize=0
 \epsfxsize=\epsftsize \epsfysize=\epsfrsize
 \epsfrsize=0pt
 \else\epsftmp=\epsftsize \divide\epsftmp\epsfrsize
 \epsfxsize=\epsfysize \multiply\epsfxsize\epsftmp
 \multiply\epsftmp\epsfrsize \advance\epsftsize-\epsftmp
 \epsftmp=\epsfysize
 \loop \advance\epsftsize\epsftsize \divide\epsftmp 2
 \ifnum\epsftmp>0
 \ifnum\epsftsize<\epsfrsize\else
 \advance\epsftsize-\epsfrsize \advance\epsfxsize\epsftmp \fi
 \repeat
 \epsfrsize=0pt
 \fi
 \else \ifnum\epsfysize=0
 \epsftmp=\epsfrsize \divide\epsftmp\epsftsize
 \epsfysize=\epsfxsize \multiply\epsfysize\epsftmp
 \multiply\epsftmp\epsftsize \advance\epsfrsize-\epsftmp
 \epsftmp=\epsfxsize
 \loop \advance\epsfrsize\epsfrsize \divide\epsftmp 2
 \ifnum\epsftmp>0
 \ifnum\epsfrsize<\epsftsize\else
 \advance\epsfrsize-\epsftsize \advance\epsfysize\epsftmp \fi
 \repeat
 \epsfrsize=0pt
 \else
 \epsfrsize=\epsfysize
 \fi
 \fi
 \ifepsfverbose\message{#1: width=\the\epsfxsize, height=\the\epsfysize}\fi
 \epsftmp=10\epsfxsize \divide\epsftmp\pspoints
 \vbox to\epsfysize{\vfil\hbox to\epsfxsize{%
 \ifnum\epsfrsize=0\relax
 \includegraphics{#1}%
 \else
 \epsfrsize=10\epsfysize \divide\epsfrsize\pspoints
 \includegraphics{#1}%
 \fi
 \hfil}}%
\global\epsfxsize=0pt\global\epsfysize=0pt}%
 {\catcode`\%=12 \global\let\epsfpercent=
\long\def\epsfaux#1#2:#3\\{\ifx#1\epsfpercent
 \def\testit{#2}\ifx\testit\epsfbblit
 \epsfgrab #3 . . . \\%
 \epsffileokfalse
 \global\epsfbbfoundtrue
 \fi\else\ifx#1\par\else\epsffileokfalse\fi\fi}%
\def\epsfempty{}%
\def\epsfgrab #1 #2 #3 #4 #5\\{%
\global\def\epsfllx{#1}\ifx\epsfllx\epsfempty
 \epsfgrab #2 #3 #4 #5 .\\\else
 \global\def\epsflly{#2}%
 \global\def\epsfurx{#3}\global\def\epsfury{#4}\fi}%
\def\epsfsize#1#2{\epsfxsize}
\pagestyle{plain}

\begin{document}

\begin{flushright}  
UdeM-GPP-TH-02-96\\
\end{flushright}
\vskip0.5truecm

\begin{center} 

{\large \bf
\centerline{T-violating Triple-Product Correlations in Charmless $\Lambda_b$ 
 Decays}}
\vspace*{1.0cm}
{\large Wafia Bensalem\footnote{email: wafia@lps.umontreal.ca},
  Alakabha Datta\footnote{email: datta@lps.umontreal.ca} 
and David
  London\footnote{email: london@lps.umontreal.ca}} \vskip0.3cm
{\it  Laboratoire Ren\'e J.-A. L\'evesque, Universit\'e de
  Montr\'eal,} \\
{\it C.P. 6128, succ.\ centre-ville, Montr\'eal, QC, Canada H3C 3J7} \\
\vskip0.5cm
\bigskip
(\today)
\vskip0.5cm
{\Large Abstract\\}
\vskip3truemm
\parbox[t]{\textwidth} {Using factorization, we compute, within the
standard model, the T-violating triple-product correlations in the
charmless decays $\Lambda_b \to F_1 F_2$, where $F_1$ is a light
spin-$1\over 2$ baryon and $F_2$ is a pseudoscalar ($P$) or vector
($V$) meson. We find a large triple-product asymmetry of 18\% for the
decay $\Lambda_b \to p K^-$. However, for other classes of $\Lambda_b
\to F_1 P$ decays, the asymmetry is found to be at most at the percent
level. For $\Lambda_b \to F_1 V$ decays, we find that all
triple-product asymmetries are small (at most $O(1\%)$) for a
transversely-polarized $V$, and are even smaller for longitudinal
polarization. Our estimates of the nonfactorizable contributions to
these decays show them to be negligible, and we describe ways of
testing this.}
\end{center}
\thispagestyle{empty}
\newpage
\setcounter{page}{1}
\textheight 23.0 true cm
\baselineskip=14pt

Over the past two decades, there has been a great deal of theoretical
work examining CP violation in the $B$ system. Most of this work has
focussed on the decays of $B$ mesons. The main reason is that the
indirect CP-violating asymmetries in $B$-meson decays can be used to
extract the interior angles of the unitarity triangle ($\alpha$,
$\beta$ and $\gamma$) with no hadronic uncertainty \cite{CPreview}.
The knowledge of these angles will allow us to test the standard model
(SM) explanation of CP violation. In order to make such measurements,
the $B$-factories BaBar and Belle have been built. These machines
produce copious numbers of $B^0$--${\bar B}^0$ pairs, and have now
provided the first definitive evidence for CP violation outside the
kaon system: $\sin 2\beta = 0.78 \pm 0.08$ \cite{betameas}.

On the other hand, in the coming years machines will be built which
are capable of producing large numbers of $\Lambda_b$ baryons. These
include hadron machines, such as the Tevatron, LHC, etc., as well as
possibly a high-luminosity $e^+ e^-$ machine running at the $Z$ pole.
People have therefore started to examine the SM predictions for a
variety of $\Lambda_b$ decays. This is a worthwhile effort, since it
is conceivable that certain types of new physics will be more easily
detectable in $\Lambda_b$ decays than in $B$ decays. For example,
$\Lambda_b$'s can be used to probe observables which depend on the
spin of the $b$-quark, whereas such observables will be unmeasurable
in $B$-meson decays.

One class of observables which may involve, among other things, the
$b$-quark spin is triple-product correlations. These take the form
$\vec v_1 \cdot (\vec v_2 \times \vec v_3)$, where each $v_i$ is a
spin or momentum. These triple products are odd under time reversal
(T) and hence, by the CPT theorem, also constitute potential signals
of CP violation. By measuring a nonzero value of the asymmetry
\beq
A_{\sss T} \equiv 
{{\Gamma (\vec v_1 \cdot (\vec v_2 \times \vec v_3)>0) - 
\Gamma (\vec v_1 \cdot (\vec v_2 \times \vec v_3)<0)} \over 
{\Gamma (\vec v_1 \cdot (\vec v_2 \times \vec v_3)>0) + 
\Gamma (\vec v_1 \cdot (\vec v_2 \times \vec v_3)<0)}} ~,
\label{Toddasym}
\eeq
where $\Gamma$ is the decay rate for the process in question, one can
establish the presence of a nonzero triple-product correlation. Note
that there is a well-known technical complication: strong phases can
produce a nonzero value of $A_{\sss T}$, even if there is no CP
violation (i.e.\ if the weak phases are zero). Thus, strictly
speaking, the asymmetry $A_{\sss T}$ is not in fact a T-violating
effect. Nevertheless, one can still obtain a true T-violating signal
by measuring a nonzero value of
\beq
{\cal A}_{\sss T} \equiv {1\over 2}(A_T-{\bar A}_T) ~,
\label{Tviolasym}
\eeq
where ${\bar A}_{\sss T}$ is the T-odd asymmetry measured in the
CP-conjugate decay process.

Recently, T-violating triple-product correlations were calculated for
the inclusive quark-level decay $b \to s \bar{u} u$ \cite{Wafia}. In
that calculation, all final-state masses were neglected. Ignoring
triple products which involve three spins, only two non-negligible
triple-product asymmetries were found. They are: (i) ${\vec p}_u \cdot
( {\vec s}_u \times {\vec s}_{\bar u} )$ or ${\vec p}_{\bar u} \cdot (
{\vec s}_u \times {\vec s}_{\bar u} )$, and (ii) $\vec s_b \cdot (\vec
p_u\times \vec p_s)$. While the former triple product can be probed in
$B \to V_1 V_2$ decays, where $V_1$ and $V_2$ are vector mesons, the
latter can only be measured in $\Lambda_b$ decays, since the spin of
the $b$ quark is involved.

In this paper we study, within the SM, the triple products in
charmless two-body $\Lambda_b$ decays which are generated by the
quark-level transitions $b \to s \bar{u} u$ or $b \to d \bar{u} u$.
These decays are of the type $\Lambda_b \to F_1 F_2$, where $F_1$ is a
light spin-$1\over 2$ baryon, such as $p$, $\Lambda$, etc., and $F_2$
is a pseudoscalar ($P$) or vector ($V$) meson. Such T-violating
triple-product correlations, along with other P-violating asymmetries,
have been studied for hyperon decays \cite{Pak}, but relatively little
work has been done to study CP violation in $\Lambda_b$ decays.

The decays $\Lambda_b \to F_1 P$ are similar to hyperon decays. As we
will see, there is a triple-product correlation in such decays of the
form ${\vec s}_{\Lambda_b} \cdot ({\vec s}_{F_1} \times {\vec p})$,
where ${\vec s}_{\Lambda_b}$ and ${\vec s}_{F_1}$ are the
polarizations of the $\Lambda_b$ and $F_1$, respectively, and ${\vec
p}$ is the momentum of one of the final-state particles in the rest
frame of the $\Lambda_b$. On the other hand, $\Lambda_b$ decays can
also include a vector meson in the final state, which is not
kinematically accesible for hyperon decays. The decay $\Lambda_b \to
F_1 V$ can give rise to a variety of triple-product correlations
involving the spin of the $\Lambda_b$ and/or $V$.

Many of these triple products involve the spin of the $\Lambda_b$.
Perhaps the easiest way to obtain this quantity is to produce the
$\Lambda_b$ baryons in the decay of an on-shell $Z$ boson. This is
because, in the decay $Z\to b{\bar b}$, the $b$-quarks have a large
average longitudinal polarization of about $-$94\%.  According to
heavy-quark effective theory, this polarization is retained when a
$b$-quark hadronizes into a $\Lambda_b$, and recent measurements of
the average longitudinal polarization of $b$-flavored baryons produced
in $Z^0$ decays (measured through their decay to $ \Lambda_c \ell
\nu_{\ell} X$) is consistent with this conclusion \cite{Opal}. Thus,
the so-called GigaZ option ($2 \times 10^9$ $Z$ bosons per year
\cite{tesla,usaLC}) of a high-luminosity $e^+ e^-$ collider running at
the $Z$ peak would be a particularly good environment for measuring
triple-product correlations in $\Lambda_b$ decays. However, even if
the spin of the $\Lambda_b$ cannot be measured at a given machine,
some of the triple-product correlations in $\Lambda_b \to F_1 V$ do
not involve the polarization of the initial state. Thus, triple
products can be measured at a variety of facilities in which a large
number of $\Lambda_b$ baryons is produced.

We begin our analysis by studying the nonleptonic decay
$\Lambda_b\to F_1 P$. The general form for this amplitude can
be written as
\beq
{\cal M}_P= A(\Lambda_b\to F_1 P) = i {\bar u}_{F_1} (a +
b\gamma_{5}) u_{\Lambda_b} ~.
\label{pscalar}
\eeq
In order to make contact with the conventional notation for hyperon
decay, we note that, in the rest frame of the parent baryon, the decay
amplitude reduces to 
\beq 
A(\Lambda_b\to F_1 P) = i \chi_{F_1} (S+P {\vec\sigma} \cdot
{\hat p}) \chi_{\Lambda_b} ~,
\eeq
where ${\hat p}$ is the unit vector along the direction of the
daughter baryon momentum, and
$S=\sqrt{2m_{\Lambda_b}(E_{F_1}+m_{F_1})}a$ and
$P=-\sqrt{2m_{\Lambda_b}(E_{F_1}-m_{F_1})}b$, where $E_{F_1}$ and
$m_{F_1}$ are, respectively, the energy and mass of the final-state
baryon $F_1$. The decay rate and the various asymmetries are given by
\beq
\Gamma=\frac{{\vec p}}{8\pi{m_{\Lambda_b}}^{2}}(|S|^2 +|P|^2) ~,~~
\alpha=\frac{2 \, {\rm Re}(S^{*}P)}{|S|^2 +|P|^2} ~,~~
\beta=\frac{2 \, {\rm Im}(S^{*}P)}{|S|^2 +|P|^2} ~,~~
\gamma=\frac{|S|^2 -|P|^2}{|S|^2 +|P|^2} ~.
\label{asymp}
\eeq
(Note: above, the quantities $\alpha$, $\beta$ and $\gamma$ should not
be confused with the CP phases of the unitarity triangle, which have
the same symbols.)

The calculation of $|{\cal M}_P|^2$ in Eq.~(\ref{pscalar}) yields
\bea
|{\cal M}_P|^2 & = & (|a|^2 - |b|^2) \left(m_{F_1} m_{\Lambda_b} +
 p_{F_1} \cdot s_{\Lambda_b} \, p_{\Lambda_b} \cdot s_{F_1} -
 p_{F_1} \cdot p_{\Lambda_b} \, s_{F_1} \cdot s_{\Lambda_b}
 \right) \nn\\
& &
+ (|a|^2 + |b|^2) \left( p_{F_1} \cdot p_{\Lambda_b} -
m_{F_1} m_{\Lambda_b} s_{F_1} \cdot s_{\Lambda_b} \right) \nn\\
& & 
+ 2 \, {\rm Re}(a b^*) \left( m_{\Lambda_b} p_{F_1} \cdot
s_{\Lambda_b} - m_{F_1} p_{\Lambda_b} \cdot s_{F_1}
\right) \nn \\ 
& & + 2 \, {\rm Im}(a b^*) \epsilon_{\mu\nu\rho\sigma}
p^\mu_{F_1} s^\nu_{F_1} p^\rho_{\Lambda_b}
s^\sigma_{\Lambda_b} ~.
\label{m2scalar}
\eea
It is the last term above which gives a triple-product correlation.
(It corresponds to $\beta$ in Eq.~(\ref{asymp}).) In the rest frame of
the $\Lambda_b$, it takes the form ${\vec p}_{F_1} \cdot ({\vec
s}_{F_1} \times {\vec s}_{\Lambda_b})$.

In order to estimate the size of this triple product, we will use
factorization to calculate ${\rm Im}(a b^*)$ at the hadron level. The
starting point is the SM effective hamiltonian for charmless hadronic
$B$ decays \cite{BuraseffH}:
\beq 
H_{eff}^q = {G_F \over \protect \sqrt{2}} [V_{ub}V^*_{uq}(c_1O_{1}^q
+ c_2 O_{2}^q) - \sum_{i=3}^{10} V_{tb}V^*_{tq} c_i^t O_i^q] + h.c.,
\label{H_eff}
\eeq
where
\bea
O_1^q = {\bar q}_\alpha \gamma_\mu L u_\beta \, {\bar u}_\beta
\gamma^\mu L b_\alpha &,& 
O_2^q = {\bar q} \gamma_\mu L u \, {\bar u} \gamma^\mu L b ~, \nn\\
O_{3(5)}^q = {\bar q} \gamma_\mu L b \, \sum_{q'} {\bar q}' \gamma^\mu L (R)
q' &,&
O_{4(6)}^q = {\bar q}_\alpha \gamma_\mu L b_\beta \, \sum_{q'} {\bar
q}'_\beta \gamma^\mu L (R) q'_\alpha ~, \\
O_{7(9)}^q = {3\over 2} {\bar q} \gamma_\mu L b \, \sum_{q'}
e_{q'} {\bar q}' \gamma^\mu R (L) q' &,&
O_{8(10)}^q = {3\over 2} {\bar q}_\alpha \gamma_\mu L b_\beta \,
\sum_{q'} e_{q'} {\bar q}'_\beta \gamma^\mu R (L) q'_\alpha ~. \nn
\label{H_effops}
\eea
In the above, $q$ can be either a $d$ or an $s$ quark, depending on
whether the decay is a $\Delta S = 0$ or a $\Delta S = -1$ process,
$q' = d$, $u$ or $s$, with $e_{q'}$ the corresponding electric charge,
and $R(L)= 1 \pm \gamma_5$. The values of the Wilson coefficients
$c_i$ evaluated at the scale $\mu = m_b = 5$ GeV, for $m_t = 176$ GeV
and $\alpha_s(m_Z) = 0.117$, are \cite{FSHe}:
\bea
c_1 = -0.324 &,& c_2 = 1.151 ~, \nonumber\\
c^t_3 = 0.017 ~,~~ c^t_4 = -0.037 &,& c^t_5 = 0.010 ~,~~
c^t_6 =-0.045 ~, \nonumber\\
c^t_7 = -1.24\times 10^{-5} ~,~~ c_8^t = 3.77\times 10^{-4} &,&
c_9^t = -0.010 ~,~~ c_{10}^t = 2.06\times 10^{-3} ~.
\eea

In our analysis we will also consider the gluonic dipole operator in
which the gluon splits into two quarks, giving the effective operator
\beq
H_{11} = i \, \frac{G_F}{\sqrt{2}} \, \frac{\alpha_s(\mu)}{2\pi k^2}
\, m_b(\mu) \, c_{11} \, V_{tb} V_{ts}^* \, \bar{s}(p_s)
\sigma_{\mu\nu} R T^a b(p_b) \bar{q}(p_2) \gamma^{\mu} T^a
q(p_1)k^{\nu} ~,
\label{H11d}
\eeq
where $k=p_b-ps$ and $c_{11} = 0.2$ \cite{Dipole}. It is often useful
to write this in the Fierz-transformed form
\beq
H_{11} = -\frac{G_F}{\sqrt{2}} \, \frac{\alpha_s(\mu)}{16\pi} \,
\frac{m_b^2(\mu)}{k^2} \, c_{11} \, \frac{N_c^2-1}{N_c^2} \, V_{tb}
V_{ts}^* \, \left[ \delta_{\alpha \beta}
\delta_{\alpha^{\prime}\beta^{\prime}} -\frac{2N_c}{N_c^2-1}
T^a_{\alpha \beta} T^a_{\alpha^{\prime} \beta^{\prime}}\right]\sum_i
T_i ~,
\eeq
where
\bea
T_1 & = & 2\bar{s}_{\alpha} \gamma_{\mu}Lq_{\beta}
\bar{q}_{\alpha^{\prime}} \gamma^{\mu}Lb_{\beta^{\prime}} -
4\bar{s}_{\alpha} Rq_{\beta}
\bar{q}_{\alpha^{\prime}}Lb_{\beta^{\prime}} ~, \nonumber\\
T_2 &= & 2\frac{m_s}{m_b}\bar{s}_{\alpha} \gamma_{\mu}Rq_{\beta}
\bar{q}_{\alpha^{\prime}} \gamma^{\mu}Rb_{\beta^{\prime}} -
4\frac{m_s}{m_b}\bar{s}_{\alpha} Lq_{\beta}
\bar{q}_{\alpha^{\prime}}Rb_{\beta^{\prime}} ~, \nonumber\\
T_3 &= & \frac{(p_b+p_s)_{\mu}}{m_b}[ \bar{s}_{\alpha}
\gamma^{\mu}Lq_{\beta} \bar{q}_{\alpha^{\prime}} Rb_{\beta^{\prime}} +
\bar{s}_{\alpha} Rq_{\beta}
\bar{q}_{\alpha^{\prime}}\gamma^{\mu}Rb_{\beta^{\prime}}] ~, \nonumber\\
T_4 &= & \frac{(p_b+p_s)_{\mu}}{m_b}[ i\bar{s}_{\alpha} \sigma^{\mu
\nu}Rq_{\beta} \bar{q}_{\alpha^{\prime}}
\gamma_{\nu}Rb_{\beta^{\prime}} - i\bar{s}_{\alpha}\gamma_{\nu}
Lq_{\beta} \bar{q}_{\alpha^{\prime}}\sigma^{\mu
\nu}Rb_{\beta^{\prime}}] ~,
\eea
in which we have defined $\sigma_{\mu \nu} =
\frac{i}{2}[\gamma_{\mu}\gamma_{\nu} -\gamma_{\nu}\gamma_{\mu}]$.

We now apply the effective hamiltonian to specific exclusive
$\Lambda_b$ decays. We will focus on those processes for which
factorization is expected to be a good approximation, namely
colour-allowed decays. We begin with $\Lambda_b \to p K^-$, which is a
$b \to s \bar{u} u$ transition. Factorization allows us to write
\beq
A(\Lambda_b\to pK^-)= \sum_{O,O'} \bra{K^-} O \ket{0} \, \bra{p} O'
\ket{\Lambda_b} ~.
\eeq
It is straightforward to show that the operators in $H_{eff}^s$ and
$H_{11}$ lead to two classes of terms in the decay amplitude: (a)
$\bra{K^-} {\bar s} \gamma^\mu (1 \pm \gamma_5) u \ket{0} \, \bra{p}
{\bar u} \gamma_\mu (1 \pm \gamma_5) b \ket{\Lambda_b}$, and (b)
$\bra{K^-} {\bar s} (1 \pm \gamma_5) u \ket{0} \, \bra{p} {\bar u} (1
\pm \gamma_5) b \ket{\Lambda_b}$. For the first of these, we define
the pseudoscalar decay constant $f_K$ as
\beq
i f_K q^{\mu}= \bra{K} \bar{s} \gamma^{\mu}(1 - \gamma_5)u \ket{0} ~,
\eeq
where $q^\mu \equiv p^\mu_{\Lambda_b} - p^\mu_p = p^\mu_K$ is the
four-momentum transfer. For the second, one can show that
\beq
\bra{K^-} {\bar s} (1 \pm \gamma_5) u \ket{0} = \mp {f_K m_K^2 \over
m_s + m_u} ~~,~~
\bra{p} {\bar u} (1 \pm \gamma_5) b \ket{\Lambda_b} = {q^\mu\over m_b} 
\bra{p} {\bar u} \gamma_\mu (1 \mp \gamma_5) b \ket{\Lambda_b} ~.
\label{LRtrans}
\eeq
(In the second matrix element, we have neglected $m_u$ compared to
$m_b$.) Thus, factorization leads to the following form for the
$\Lambda_b \to p K^-$ amplitude:
\beq
A(\Lambda_b\to pK^-)= if_K q^{\mu} \bra{p} \bar{u}
\gamma_\mu(1-\gamma_5) b\ket{\Lambda_b} X_K + i f_K q^{\mu}
\bra{p}\bar{u} \gamma_\mu(1+\gamma_5) b\ket{\Lambda_b} Y_K ~.
\label{ampdef}
\eeq 

Like any CP-violating observable, a nonzero triple product can arise
only if there are two interfering amplitudes. This will occur only if
both $X_K$ and $Y_K$ are nonzero. Since all the operators
$O_1$--$O_{10}$ involve a left-handed $b$-quark, it is clear that $X_K
\ne 0$ in the SM. Furthermore, though it is less obvious, one can also
have $Y_K \ne 0$. Consider, for example, the operator $O_6$ of
Eq.~(\ref{H_effops}). After performing Fierz transformations, this can
be written as
\beq
O_6 \sim {\bar s} (1 + \gamma_5) u \, {\bar u} (1 - \gamma_5) b ~.
\label{O6}
\eeq
However, according to Eq.~(\ref{LRtrans}), $\bra{p} {\bar u} (1 -
\gamma_5) b \ket{\Lambda_b}$ can be related to $\bra{p} {\bar u}
\gamma_\mu (1 + \gamma_5) b \ket{\Lambda_b}$. Thus, $Y_K$ receives
contributions from operators such as $O_6$. We find
\bea
X_K & = & \frac{G_F}{\sqrt{2}} \left[ V_{ub}V_{us}^* a_2-
\sum_{q=u,c,t}V_{qb}V_{qs}^* (a_4^q +a_{10}^q) - V_{tb}V_{ts}^* a_d
\left( 1+\frac{2E_K}{m_b} \right) \right] ~, \nonumber\\
Y_K & = & -\frac{G_F}{\sqrt{2}} \left[ \sum_{q=u,c,t}V_{qb}V_{qs}^*
(a_6^q +a_8^q)+\frac{5}{4} V_{tb} V_{ts}^* a_d \right] \chi_K ~,
\label{pk}
\eea
with
\beq
\chi_K = \frac{2 m_K^2}{(m_s+m_u)m_b} ~~,~~~~
a_d = \frac{\alpha_s(\mu)}{16\pi} \langle \frac{m_b^2(\mu)}{k^2}
\rangle c_{11} \frac{N_c^2-1}{N_c^2} ~.
\label{pk2}
\eeq
In the above, we have defined $a_i^q=c_i^q+ \frac{c_{i+1}^q}{N_c}$ for
$i$ odd and $a_i^q=c_i^q+ \frac{c_{i-1}^q}{N_c}$ for $i$ even. We
estimate the average gluon momentum in the dipole operator to be
$\langle m_b^2/k^2 \rangle = \int \phi_K(x)m_b^2/k^2 dx$, where the
gluon momentum in the heavy-quark limit is $k^2=m_b^2(1-x)$ and
$\phi_K$ is the kaon light-cone distribution. Choosing the asymptotic
form $\phi_K=6x(1-x)$, we find $\langle m_b^2/k^2 \rangle = 3$, which
leads to $a_d = 0.0021$.

Now, the vector and axial-vector matrix elements between the
$\Lambda_b$ and $p$ can be written in the general form
\begin{eqnarray}
\bra{p} \bar{u} \gamma^\mu b \ket{\Lambda_b} & =& \bar{u}_{p} \left[
f_1 \gamma^\mu + i \frac{f_2}{m_{\Lambda_b}}\sigma^{\mu\nu} q_\nu +
\frac{f_3}{m_{\Lambda_b}} q^\mu \right] u_{\Lambda_b} \nonumber \\
\bra{p} \bar{u} \gamma^\mu \gamma_5 b \ket{\Lambda_b} & =& \bar{u}_{p}
\left[ g_1 \gamma^\mu + i \frac{g_2}{m_{\Lambda_b}}\sigma^{\mu\nu}
q_\nu + \frac{g_3}{m_{\Lambda_b}} q^\mu \right] \gamma_5 u_{\Lambda_b}
~,
\label{genamps}
\end{eqnarray}
where the $f_i$ and $g_i$ are Lorentz-invariant form
factors. Heavy-quark symmetry imposes constraints on these form
factors. A systematic expansion of these form factors, including
$1/m_b$ corrections, has been calculated \cite{Dattaff}: in the
$m_b\to \infty$ limit, one obtains the relations
\beq
f_1 = g_1 ~~,~~~~ f_2 = g_2 = f_3 = g_3 ~.
\label{hqet}
\eeq
Using the above expressions, we find that the parameters $a$ and $b$
of Eq.~(\ref{pscalar}) can be written as
\bea
a_K & = & f_K (X_K+Y_K)\left[(m_{\Lambda_b}-m_p)f_1
+f_3\frac{m_K^2}{m_{\Lambda_b}} \right] ~, \nonumber\\
b_K & = & f_K (X_K-Y_K)\left[(m_{\Lambda_b}+m_p)g_1
-g_3\frac{m_K^2}{m_{\Lambda_b}} \right] ~.
\label{abTP}
\eea

According to Eq.~\ref{m2scalar}, the triple product in $\Lambda_b\to
pK^-$ is proportional to ${\rm Im}(a_K b_K^*)$, which is in turn
proportional to ${\rm Im}(X_K Y_K^*)$. Since $X_K$ and $Y_K$ are both
nonzero, and have different weak phases [Eq.~(\ref{pk})], we expect a
nonzero triple-product asymmetry in $\Lambda_b\to pK^-$ of the form
${\vec p}_{p} \cdot ({\vec s}_{p} \times {\vec s}_{\Lambda_b})$. At
first sight, this appears to contradict the results of
Ref.~\cite{Wafia}, since no triple products involving two spins were
found in the quark-level decay $b \to s \bar{u} u$. However, note that
$Y_K$ is proportional to $\chi_K$, which is formally suppressed by
$1/m_b$.  Thus, in the limit $m_b \to \infty$, one has $Y_K=0$, so
that the triple-product correlation will vanish. This agrees with the
conclusions of Ref.~\cite{Wafia}, which neglects the masses of the
final-state quarks (i.e.\ the limit $m_b \to \infty$ is implicitly
assumed).

However, the key point is that, for finite $m_b$, $\chi_K$ is not
small because of the presence of the chiral enhancement term
$m_K^2/(m_s+m_u)$. In fact, for $m_s=100$ MeV and $m_b= 5$ GeV,
$\chi_K \sim 1$, and hence is clearly non-negligible. The
triple-product asymmetry of ${\vec p}_{p} \cdot ({\vec s}_{p} \times
{\vec s}_{\Lambda_b})$ may therefore be sizeable. Note that this
triple product requires the measurement of both the $\Lambda_b$ and
the $p$ polarizations. If the measurement of the proton polarization
is not possible, one can instead consider a final state with an
excited nucleon, such as $\Lambda_b \to N(1440) K^-$. In this case the
polarization of the $N(1440)$ can be determined from its decay
products. (Alternatively, one can consider the decay $\Xi_b \to
\Sigma^+ K^-$, where $\Xi_b$ has quark content $bus$.)

Note also that in Eq.~(\ref{pk}) we have included the up- and
charm-quark penguin pieces, proportional to $V_{ub}V_{us}^*$ and
$V_{cb}V_{cs}^*$ respectively. These are generated by rescattering of
the tree-level operators in the effective Hamiltonian in
Eq.~(\ref{H_eff}). As we will see, the contributions from these
rescattering terms are very important. The coefficients associated
with these terms are given by
\bea
c_{3,5}^i = -c_{4,6}^i/N_c = P^i_s/N_c &,&
c_{7,9}^i = P^i_e ~,~~
c_{8,10}^i = 0 ~,~~
i=u,c ~,
\label{coeffs}
\eea
where $N_c$ is the number of colours. The leading contributions to
$P^i_{s,e}$ are given by $P^i_s = ({\frac{\alpha_s}{8\pi}}) c_2
({\frac{10}{9}} +G(m_i,\mu,q^2))$ and $P^i_e =
({\frac{\alpha_{em}}{9\pi}}) (N_c c_1+ c_2) ({\frac{10}{9}} +
G(m_i,\mu,q^2))$, in which the function $G(m,\mu,q^2)$ takes the form
\begin{eqnarray}
G(m,\mu,q^2) = 4\int^1_0 x(1-x) \mbox{ln}{m^2-x(1-x)q^2\over
\mu^2} ~\mbox{d}x ~,
\label{rescatt}
\end{eqnarray}
where $q$ is the momentum carried by the virtual gluon in the penguin
diagram. Of course, we are really interested in the matrix elements
of the various operators for the decay $\Lambda_b \to p K^-$, and so
the coefficients in Eq.~(\ref{coeffs}) should be understood to be
\beq
{\bar{c}}_i^{u,c} = \frac{\bra{p K^-}c_i^{u,c}(q^2)O_i\ket{\Lambda_b}}
{\bra{p K^-}O_i\ket{\Lambda_b}} ~.
\eeq
We will henceforth drop the distinction between ${\bar{c}}_i^{u,c}$
and $c_i^{u,c}$, with the understanding that it is the
${\bar{c}}_i^{u,c}$ which appear in the amplitude.

The analysis of other colour-allowed $\Lambda_b$ decays follows
straightforwardly from that for $\Lambda_b \to p K^-$. For example,
consider $\Lambda_b \to p \pi^-$, which is generated by the
quark-level decay $b \to d \bar{u} u$. The amplitude for $\Lambda_b
\to p \pi^-$ is given by Eq.~(\ref{pscalar}), with
\bea
a_\pi & = & f_{\pi} (X_\pi+Y_\pi)\left[(m_{\Lambda_b}-m_p)f_1
+f_3\frac{m_\pi^2}{m_{\Lambda_b}} \right] ~, \nonumber\\
b_\pi & = & f_{\pi} (X_\pi-Y_\pi)\left[(m_{\Lambda_b}+m_p)g_1
-g_3\frac{m_\pi^2}{m_{\Lambda_b}} \right] ~,
\eea
where
\bea
X_\pi & = & \frac{G_F}{\sqrt{2}} \left[ V_{ub}V_{ud}^* a_2-
\sum_{q=u,c,t}V_{qb}V_{qd}^* (a_4^q +a_{10}^q) - V_{tb}V_{td}^*
a_d(1+\frac{2E_{\pi}}{m_b}) \right] ~, \nonumber\\
Y_\pi & = & -\frac{G_F}{\sqrt{2}} \left[ \sum_{q=u,c,t}V_{qb}V_{qd}^*
(a_6^q +a_8^q)+\frac{5}{4} V_{tb} V_{td}^* a_d \right] \chi_{\pi} ~,
\label{ppi}
\eea
with
\beq
\chi_{\pi} = \frac{2 m_{\pi}^2}{(m_d+m_u)m_b} ~.
\eeq

Finally, we consider the decay $\Lambda_b \to \Lambda \eta
(\eta^{\prime})$ \cite{Dattaeta}, which is dominated by a
colour-allowed $b\to s$ penguin transition (there is also a small
colour-suppressed tree contribution). For the decay $\Lambda_b \to
\Lambda \eta$ we get
\bea
a_\eta & = & f_\pi (X_\eta+Y_\eta)\left[(m_{\Lambda_b}-m_{\Lambda})f_1
+f_3\frac{m_{\eta}^2}{m_{\Lambda_b}} \right] ~, \nonumber\\
b_\eta & = & f_\pi (X_\eta-Y_\eta)\left[(m_{\Lambda_b}+m_{\Lambda})g_1 
-g_3\frac{m_{\eta}^2}{m_{\Lambda_b}}
\right] ~,
\eea
where
\bea
X_\eta & = & \frac{G_F}{\sqrt{2}} \left[ V_{ub}V_{us}^* a_1r_1-
\sum_{q=u,c,t}V_{qb}V_{qs}^* (r_1A_q +r_2B_q) - V_{tb}V_{ts}^* r_2 a_d
\left( 1+\frac{2E_{\eta}}{m_b} \right) \right] ~, \nonumber\\
Y_\eta & = & -\frac{G_F}{\sqrt{2}} \left[ \sum_{q=u,c,t}V_{qb}V_{qs}^*
(a_6^q -\frac{1}{2}a_8^q)+\frac{5}{4} V_{tb} V_{ts}^* a_d \right]
r_2 \chi_{\eta} ~,
\label{leta}
\eea
with
\bea
A_q & = & 2a_3^q-2a_5^q-\frac{1}{2}a_7^q +\frac{1}{2}a_9^q ~,
\nonumber\\
B_q & = & a_3^q+a_4^q-a_5^q+\frac{1}{2}a_7^q -\frac{1}{2}a_9^q
-\frac{1}{2}a_{10}^q ~, \nonumber\\
\chi_{\eta} & = & \frac{m_{\eta}^2}{m_s m_b} ~.
\eea 
In the above, we have defined $r_1=f_{\eta}^u/f_{\pi}$ and
$r_2=f_{\eta}^s/f_{\pi}$, with
\bea
if_{\eta}^u p_{\eta}^{\mu} & = &
\bra{\eta}\bar{u}\gamma^{\mu}(1-\gamma_5)u\ket{0}
=\bra{\eta}\bar{d}\gamma^{\mu}(1-\gamma_5)d\ket{0} ~, \nonumber\\
if_{\eta}^sp_{\eta}^{\mu} & = &
\bra{\eta}\bar{s}\gamma^{\mu}(1-\gamma_5)s\ket{0} ~.
\eea
The amplitude for $\Lambda_b \to \Lambda \eta^{\prime}$ has the same
form as Eq.~(\ref{leta}) with the replacement $\eta \to
\eta^{\prime}$. Note that the polarization of the final-state
$\Lambda$ can be measured via its decay $\Lambda \to p \pi^{-}$.

The above analysis has been performed within the framework of
factorization. Before turning to estimates of the size of the
triple-product asymmetries, it is useful at this point to address the
issue of nonfactorizable corrections. Nonfactorizable effects are
known to be important for hyperon and charmed-baryon nonleptonic
decays, but are expected to be negligible for non-leptonic $\Lambda_b$
decays. An unambiguous signal for the presence of nonfactorizable
effects would be the observation of the decay $\Lambda_b \to \Delta^+
K^- (\pi^-)$, $\Lambda_b \to \Sigma \eta (\eta^\prime)$, or $\Lambda_b
\to \Sigma \phi$. This is because, for the factorizable contribution,
the light diquark in the $\Lambda_b$ baryon remains inert during the
weak decay. Thus, since the light diquark is an isosinglet, and since
strong interactions conserve isospin to a very good approximation, the
above $\Lambda_b$ decays are forbidden within factorization
\cite{Dattalip}.

One way to estimate the size of nonfactorizable corrections is by
using the pole model. In this model, one assumes that the
nonfactorizable decay amplitude receives contributions primarily from
one-particle intermediate states, and that these contributions then
show up as simple poles in the decay amplitude. An example of
intermediate single-particle states is the ground-state
positive-parity baryons. Consider the decay $\Lambda_b \to p K^-$. One
nonfactorizable contribution is described by the diagram in which
there is a $\Lambda_b \to \Sigma^0$ weak transition through a $W$
exchange, followed by the strong decay $\Sigma^0 \to p K^-$. The pole
contribution to the parity-violating amplitude, $a$, in
Eq.~(\ref{pscalar}) is known to be small for charmed-baryon decays
\cite{Cheng1}, and we assume this to be the case here as well. For the
parity-conserving amplitude, $b$, in Eq.~(\ref{pscalar}), we can then
write
\beq
b_{nonfac} \sim V_{ub}V_{us}^* \frac{\bra{\Sigma^0}H_w\ket{\Lambda_b}}
{m_{\Lambda_b}-m_{\Sigma_0}} g_{\Sigma^0 pK^{-}} ~,
\eeq
where $g_{\Sigma^0 pK^{-}}$ is the strong-coupling vertex which will
depend on the energy of the emitted kaon. We can use heavy-quark and
flavour $SU(3)$ symmetry to set $\bra{\Sigma^0}H_w\ket{\Lambda_b} \sim
\bra{\Sigma^+}H_w\ket{\Lambda_c}$. Writing the weak matrix element
$\bra{\Sigma^+}H_w\ket{\Lambda_c}= \frac{G_F}{\sqrt{2}}m^3$, we obtain
\beq
{b_{nonfac} \over b_{fac}} \sim \frac{m}{f_K}\frac{m^2}
{(m_{\Lambda_b}-m_{\Sigma^0})(m_{\Lambda_b}+m_p)}g_{\Sigma^0 pK^{-}}
~,
\label{nf}
\eeq
where we have chosen the tree-level term for $A_{fac}$. Since the
emitted kaon is hard and since the quarks inside it are energetic, the
strong coupling $g_{\Sigma^0 p K^-}\sim \alpha_s(\mu \sim E_K \sim
m_b)$. In other words, the offshell $\Sigma^0$ has to emit a hard
gluon to create a $\bar{u} u$ pair to form the $p K^-$ final
state. The matrix element $m$ can either be estimated using a model
\cite{Cheng1}, or obtained from a fit to the charmed baryon decay
$\Lambda_c \to \Sigma^0 \pi^+$ \cite{Dattanl}. In both cases one
obtains $m \sim 0.1-0.2$ GeV, so that, from Eq.~(\ref{nf}), the
nonfactorizable corrections are found to be tiny. Arguments for small
nonfactorizable effects in $\Lambda_b$ decays can also be made based
on the total width calculations \cite{Cheng2}.

To summarize: for colour-allowed $\Lambda_b\to F_1 P$ decays, we find
that the triple-product correlation ${\rm Im}(a b^*) {\vec p}_{F_1}
\cdot ({\vec s}_{F_1} \times {\vec s}_{\Lambda_b})$ can be nonzero.
The next step is to calculate the size of the asymmetry ${\cal
A}_{\sss T}$ in Eq.~(\ref{Tviolasym}) for the various decays.

We begin with $\Lambda_b \to p K^-$. For this decay, we use the
expressions for $a_K$ and $b_K$ found in Eq.~(\ref{abTP}). We note
that the $f_3$ ($g_3$) term is suppressed relative to the $f_1$
($g_1$) term by a factor $m_K^2/m_{\Lambda_b}^2 \sim 0.01$, and so can
be neglected (and similarly for the $g_3$ piece). Furthermore, we take
$f_1 = g_1$ [Eq.~(\ref{hqet})], in which case all dependence on this
form factor cancels in $A_{\sss T}$ [Eq.~(\ref{Toddasym})]. The
quantities $a_K$ and $b_K$ depend on the parameters of the
Cabibbo-Kobayashi-Maskawa (CKM) matrix, whose values are taken to be
\beq
\rho = 0.17 ~~,~~~~ \eta = 0.39 ~.
\eeq
For the chiral enhancement term $\chi_K$ [Eq.~(\ref{pk2})], we
take $\chi_K = 1$.

In order to estimate the value of rescattering terms, one has to
choose a value of $q^2$. We consider two possibilities:
\beq
{\rm Model~1:}~~q^2 = {m_b^2 \over 4} ~~,~~~~ 
{\rm Model~2:}~~q^2 = {m_b^2 \over 2} ~.
\eeq
Taking $m_c = 1.4$ GeV, $m_u = 6$ MeV and $m_b = 5$ GeV, and writing
$c_4^c = |c_4^c| e^{i \delta^c}$, we find
\bea
& {\rm Model~1:} & |c_4^u| = 0.02 ~~,~~ |c_4^c| = 0.02 ~~,~~ \delta^c
= 51^\circ ~, \nn\\
& {\rm Model~2:} & |c_4^u| = 0.021 ~~,~~ |c_4^c| = 0.015 ~~,~~ \delta^c
= 0 ~.
\eea
(In accordance with CPT, we set the phase of $c_4^u$ to zero
\cite{Hou}.)

Before presenting the numerical analysis, it is useful to anticipate
the results. Referring to Eqs.~(\ref{m2scalar}), (\ref{abTP}) and
(\ref{pk}), we expect the triple-product asymmetry to be of order
\beq
{2 {\rm Im} (a_Kb_K^*) \over |a_K|^2 + |b_K|^2 + 2 {\rm Re}(a_Kb_K^*)}
\simeq {{\rm Im}(X_K Y_K^*) \over |X_K|^2} \simeq {a_2 a_6 \eta
\lambda^2 \over a_2^2 \lambda^4 + a_4^2} = 24\% ~.
\label{Pestimate}
\eeq
Of course, this is a back-of-the-envelope estimate, but it does
indicate that we can expect a reasonably large asymmetry, even when
the rescattering effects are included. The fundamental reason for this
is the following: the triple product is due mainly to the interference
of the $V_{ub}V_{us}^*$ piece of $X_K$ (we refer to this as $T$, the
``tree'') and the $V_{tb}V_{ts}^*$ piece of $Y_K$ ($P$, the
``penguin''). Like any CP-violating quantity, the asymmetry will
therefore be maximized when the two interfering amplitudes are of
comparable size. A quick calculation of these two quantities in Model
2 above shows that $|T/P| = 0.35$. The two amplitudes are therefore
similar in size, leading to the sizeable asymmetry estimate above.

We have performed the phase-space integration for $\Lambda_b \to p
K^-$ using the computer program RAMBO. For Model 1, we find that
$A_{\sss T} = -20.8\%$ and ${\bar A}_{\sss T} = +15.0\%$, leading to a
T-violating asymmetry of ${\cal A}_{\sss T}^{pK} = -17.9\%$. In Model
2, since the strong phase vanishes, one necessarily has ${\bar
A}_{\sss T} = -A_{\sss T}$, and we find ${\cal A}_{\sss T}^{pK} =
-19.1\%$. (We note in passing that the rescattering effects are quite
important. Without them, the asymmetry would be ${\cal A}_{\sss
T}^{pK} = -26.1\%$. Thus, their inclusion leads to a correction in the
asymmetry of about 25\%.) These numbers are all consistent with the
estimate in Eq.~(\ref{Pestimate}). We therefore conclude that the SM
predicts a sizeable triple-product asymmetry in the decay $\Lambda_b
\to p K^-$. (Note that, since the estimate in Eq.~(\ref{Pestimate})
uses only the values of the Wilson coefficients and the CKM matrix
elements, we expect a large asymmetry even if nonfactorizable
contributions are present.)

There is one digressionary remark which is worth making here. From the
measurement of $\epsilon_K$, the CP-violating parameter in the kaon
system, we know that the product $B_K \eta$ is positive, where $B_K$
is the kaon bag parameter and $\eta$ is the CP-violating CKM
parameter. It is usually assumed that $B_K > 0$, so that $\eta$ is
also positive, and the unitarity triangle points up. However, there is
no experimental evidence yet that $B_K > 0$. The T-violating
triple-product asymmetry in $\Lambda_b \to p K^-$ is proportional to
$\eta \cos(\delta)$, where $\delta$ is a strong phase. If one assumes
that $|\delta| < 90^\circ$, which is strongly favoured theoretically,
then the triple product asymmetry measures the sign of $\eta$. This
provides a cross check to the information obtained from the kaon
system.

Turning now to the decay $\Lambda_b \to p \pi^-$, we have applied this
same analysis as above. Taking $m_d = m_u = 6$ MeV, we have $\chi_\pi
= 0.65$. In this case, the tree amplitude $T$ is larger than the
penguin amplitude $P$, with $|P/T| = 0.08$. Because these two
interfering amplitudes are less comparable in size than was the case
for $\Lambda_b \to p K^-$, we expect a correspondingly smaller
asymmetry. This is indeed what is found. In Model 1, we have $A_{\sss
T} = 6.3\%$ and ${\bar A}_{\sss T} = -4.5\%$, so that ${\cal A}_{\sss
T}^{p\pi} = 5.4\%$. Model 2 gives a similar asymmetry: ${\cal A}_{\sss
T}^{p\pi} = 5.6\%$.

For the decays $\Lambda_b \to \Lambda \eta$ and $\Lambda_b \to \Lambda
\eta'$, we have to define the quark content and mixing of the physical
$\eta$ and $\eta'$ mesons. We use the Isgur mixing \cite{Isgur}:
\beq
\bra{\eta} = \frac{1}{\sqrt{2}}[N-S] ~~,~~~~
\bra{\eta^{\prime}}= \frac{1}{\sqrt{2}}[N+S] ~,
\eeq
where $N= [\bra{u\bar{u}} + \bra{d\bar{d}}]/\sqrt{2}$ and $S=
\bra{s\bar{s}}$. $SU(3)$ symmetry then gives
\beq
f_{\eta}^u=f_{\pi}/2 ~~,~~~~
f_{\eta}^s=-f_{\pi}/\sqrt{2} ~~,~~~~
f_{\eta^{\prime}}^u=f_{\pi}/2 ~~,~~~~
f_{\eta^{\prime}}^s=f_{\pi}/\sqrt{2} ~,
\eeq
where $f_{\pi}=131$ MeV. We also take $\chi_\eta = 0.6$ and
$\chi_{\eta'} = 1.8$. For both decays the interfering amplitudes are
very different in size: $|T/P| = 0.03$ and 0.01 for the $\Lambda\eta$
and $\Lambda\eta'$ final states, respectively. We can therefore expect
to obtain tiny triple-product asymmetries, and this should hold even
if nonfactorizable effects are present. For $\Lambda_b \to \Lambda
\eta$ we have ${\cal A}_{\sss T}^{\Lambda\eta} = 0.6\%$ (Model 1) or
0.9\% (Model 2), while for $\Lambda_b \to \Lambda \eta'$, ${\cal
A}_{\sss T}^{\Lambda\eta'} = -0.6\%$ (Model 1) or $-0.5\%$ (Model
2). It is unlikely that such tiny asymmetries can be
measured. However, this also suggests that these processes might be
good areas to search for new physics \cite{newpaper}.

We now turn to the decays $\Lambda_b\to F_1 V$. The general decay
amplitude can be written as \cite{SPT}
\beq
{\cal M}_V= Amp(\Lambda_{F_1}\to B V) = {\bar u}_{F_1}
\varepsilon^*_\mu \left[ (p_{\Lambda_b}^\mu + p_{F_1}^\mu)
(a+b\gamma_{5}) + \gamma^{\mu} (x+y\gamma_{5}) \right] u_{\Lambda_b}
~,
\label{vector}
\eeq
where $\varepsilon^*_\mu$ is the polarization of the vector meson. In
the rest frame of the $\Lambda_b$, we can write $p_V =
(E_V,0,0,|\vec{p}|)$ and $p_{F_1} = (E_{F_1},0,0,-|\vec{p}|)$. Thus,
it is clear that $\varepsilon^*_V \cdot (p_{\Lambda_b} + p_{F_1})$
will be nonzero only for a longitudinally-polarized $V$. This will be
important in what follows.

The calculation of $|{\cal M}_V|^2$ gives the following triple-product
terms:
\bea
|{\cal M}_V|^2_{t.p.} & = & 2 \, {\rm Im}(a b^*) \left\vert
\varepsilon_V \cdot (p_{\Lambda_b} + p_{F_1}) \right\vert^2 \,
\epsilon_{\mu\nu\rho\sigma} p^\mu_{F_1} s^\nu_{F_1} p^\rho_{\Lambda_b}
s^\sigma_{\Lambda_b} \nn\\
&& + 2 \, {\rm Im} \left(x y^*\right) \epsilon_{\alpha\beta\mu\nu}
\left[ \varepsilon_V \cdot s_{F_1} p_{F_1}^\alpha p_{\Lambda_b}^\beta
s_{\Lambda_b}^\mu \varepsilon_V^\nu - \varepsilon_V \cdot p_{F_1}
s_{F_1}^\alpha p_{\Lambda_b}^\beta s_{\Lambda_b}^\mu \varepsilon_V^\nu
\right. \nn\\
&& \hskip1.2truein \left. + \varepsilon_V \cdot s_{\Lambda_b}
p_{F_1}^\alpha s_{F_1}^\beta \varepsilon_V^\mu p_{\Lambda_b}^\nu -
\varepsilon_V \cdot p_{\Lambda_b} p_{F_1}^\alpha s_{F_1}^\beta
\varepsilon_V^\mu s_{\Lambda_b}^\nu \right] \nn\\
&& + 2 \, \varepsilon_V \cdot \left( p_{\Lambda_b} + p_{F_1} \right)
\epsilon_{\alpha\beta\mu\nu} \left[ {\rm Im} \left( a x^* + b y^*
\right) p_{F_1}^\alpha s_{F_1}^\beta p_{\Lambda_b}^\mu
\varepsilon_V^\nu \right. \nn\\
&& \hskip1.8truein + m_{\Lambda_b} {\rm Im} \left( b x^* + a y^*
\right) p_{F_1}^\alpha s_{F_1}^\beta s_{\Lambda_b}^\mu
\varepsilon_V^\nu \nn\\
&& \hskip1.8truein - {\rm Im} \left( a x^* - b y^* \right)
p_{F_1}^\alpha p_{\Lambda_b}^\beta s_{\Lambda_b}^\mu \varepsilon_V^\nu
\nn\\ && \hskip1.8truein \left. - m_{F_1} {\rm Im} \left( a y^* - b
x^* \right) s_{F_1}^\alpha p_{\Lambda_b}^\beta s_{\Lambda_b}^\mu
\varepsilon_V^\nu \right] ~.
\label{vecTPs}
\eea
Note that if we sum over the polarization of the vector meson, we
essentially reproduce the results found for $\Lambda_b\to F_1 P$. That
is, there is only one triple product, which takes the form
$\epsilon_{\mu\nu\rho\sigma} p^\mu_{F_1} s^\nu_{F_1}
p^\rho_{\Lambda_b} s^\sigma_{\Lambda_b}$.

As usual, we use factorization to calculate the coefficients $a$, $b$,
$x$ and $y$. Consider first the decay $\Lambda_b \to p K^{*-}$. We
define the decay constant ${g_{K^*}}$ as
\beq
m_{K^*}g_{K^*}\varepsilon_{\mu}^{*} = \bra{K^*}
\bar{s}\gamma_{\mu}u\ket{0} ~.
\label{K*vector}
\eeq
In general, factorization allows us to write
\bea
A(\Lambda_b\to pK^{*-}) &=& m_{K^*} g_{K^*} \left\{
\varepsilon_{\mu}^{*} \bra{p} \bar{u} \gamma^\mu(1-\gamma_5)
b\ket{\Lambda_b} X_{K^*} \right. \nn\\
&& \qquad\qquad + \, \varepsilon_{\mu}^{*} \bra{p}\bar{u}
\gamma^\mu(1+\gamma_5) b\ket{\Lambda_b} Y_{K^*} \\
&& \qquad\qquad + \, \varepsilon \cdot (p_{\Lambda_b} + p_p) \, q_\mu \,
\bra{p}\bar{u} \gamma^\mu(1-\gamma_5) b\ket{\Lambda_b} A_{K^*} \nn\\
&& \qquad\qquad \left.  + \, \varepsilon \cdot (p_{\Lambda_b} + p_p) \,
q_\mu \, \bra{p}\bar{u} \gamma^\mu(1+\gamma_5) b\ket{\Lambda_b}
B_{K^*} \right\}~. \nn
\eea
The coefficients $X_{K^*}$, $Y_{K^*}$, $A_{K^*}$ and $B_{K^*}$ can be
calculated using the effective hamiltonian. As noted earlier, $A_{K^*}$
and $B_{K^*}$ are nonzero only for a longitudinally-polarized
$K^{*-}$.

Consider first the operators $O_1$--$O_{10}$. Since all of these lead
to $K^{*-}$ matrix elements of the form in Eq.~(\ref{K*vector}), none
of them can contribute to $A_{K^*}$ and $B_{K^*}$. Furthermore, one
can show that none of these give $Y_{K^*} \ne 0$ either. For example,
consider again the operator $O_6$, which led to $Y_K \ne 0$. Because
$\bra{K^{*-}} {\bar s} (1 + \gamma_5) u \ket{0} = 0$, $O_6$ will not
contribute to $Y_{K^*}$. Thus, within factorization, if we restrict
ourselves only to the operators $O_1$--$O_{10}$, the only nonzero
coefficient is $X_{K^*}$, which means that all triple products vanish,
since there is only a single decay amplitude.

In order to generate triple products in $\Lambda_b \to p K^{*-}$, it
is necessary to consider the dipole operator $O_{11}$, whose effective
coefficient is rather small (Eq.~(\ref{pk2}): $a_d = 0.0021$).
However, there is an important observation one can make. The
contributions of $O_{11}$ to $Y_{K^*}$, $A_{K^*}$ and $B_{K^*}$ all
involve the tensor matrix element for $K^{*-}$, which we define as
\beq
-ig_{K^*}^T\left[\varepsilon_{\mu}^{*}p_{\nu}^{K^*}-
  \varepsilon_{\nu}^{*}p_{\mu}^{K^*}\right] = \bra{K^*}
\bar{s}\sigma_{\mu \nu}u\ket{0} ~.
\label{K*tensor}
\eeq
Now, in the rest frame of the $\Lambda_b$, we can write
$p_{K^{*}}=(E_{K^*},0,0,|\vec{p}_{K^*}|)$. In the heavy-quark limit,
in which $E_{K^*} \gg m_{K^*}$, the longitudinal polarization vector
can be written approximately as
\beq
\varepsilon_\mu^{\lambda=0} \simeq {1 \over m_{K^*}} \left(
p_{K^*}^\mu + {m_{K^*}^2 \over 2 E_{K^*}} n^\mu \right),
\eeq
with $n^\mu = (-1,0,0,1)$. From Eq.~(\ref{K*tensor}), we see that the
piece of $\varepsilon_\mu^{\lambda=0}$ which is proportional to
$p_{K^*}^\mu$ will not contribute to the matrix element. Thus, in the
heavy-quark limit, we have $A_{K^*} \simeq B_{K^*} \approx 0$.
Furthermore, we expect that the value of $Y_{K^*}$ for a
longitudinally-polarized $K^{*-}$ meson will be suppressed relative to
that for a transversely-polarized $K^{*-}$ by about $m_{K^*} / 2
E_{K^*} = 16\%$. As we will see, $Y_{K^*}$ is already small for a
transversely-polarized $K^{*-}$, so that $Y_{K^*} \approx 0$ for
longitudinal polarization. Therefore, any triple products in
$\Lambda_b \to p K^{*-}$ should be largest for a
transversely-polarized $K^{*-}$, although we expect even these to be
small.

Considering separately the longitudinal ($\lambda = 0$) and transverse
($\lambda = \perp$) polarizations of the final-state vector meson, we
find
\bea
X_{K^*}^{\lambda=\perp} & = & \frac{G_F}{\sqrt{2}} \left[ V_{ub}V_{us}^*
a_2- \sum_{q=u,c,t}V_{qb}V_{qs}^* (a_4^q +a_{10}^q) - V_{tb}V_{ts}^*
\frac{a_d}{2} \right] ~, \nonumber\\
X_{K^*}^{\lambda=0} & = & \frac{G_F}{\sqrt{2}} \left[ V_{ub}V_{us}^* a_2-
\sum_{q=u,c,t}V_{qb}V_{qs}^* (a_4^q +a_{10}^q) - V_{tb}V_{ts}^* a_d
\left( 1+\frac{2E_{K^*}}{m_b} \right) \right] ~, \nonumber\\
Y_{K^*}^{\lambda=\perp} & = & \frac{G_F}{\sqrt{2}}V_{tb}V_{ts}^* a_d ~, \nn\\
Y_{K^*}^{\lambda=0} & \approx & 0 ~,
\label{pkstar}
\eea
where
\beq
z \equiv \frac{E_{K^*}}{m_{K^*}}\frac{g_{K^*}^T}{g_{K^*}} ~.
\eeq
For $g_{K^*} = 226$ MeV and $g_{K^*}^T = 160$ MeV \cite{Ball}, $z =
2.23$. To a good approximation, the quantities $a$, $b$, $x$ and $y$
of Eq.~(\ref{vector}) can then be expressed as
\bea
a_{K^*}^\lambda &=& m_{K^{*}} g_{K^{*}}\frac{f_{2}}{m_{\Lambda_b}}
[X_{K^*}^{\lambda}+ z Y_{K^*}^{\lambda}] ~, \nn\\
b_{K^*}^\lambda &=& -m_{K^{*}} g_{K^{*}}\frac{g_{2}}{m_{\Lambda_b}}
[X_{K^*}^{\lambda}- z Y_{K^*}^{\lambda}] ~, \nn\\
x_{K^*}^\lambda &=& m_{K^{*}} g_{K^{*}} [f_{1}-
\frac{m_p+m_{\Lambda_b}}{m_{\Lambda_b}}f_{2}][X_{K^*}^{\lambda}+ z
Y_{K^*}^{\lambda}] ~, \nonumber\\
y_{K^*}^\lambda &=& -m_{K^{*}} g_{K^{*}} [g_{1}+
\frac{m_{\Lambda_b}-m_p}{m_{\Lambda_b}}g_{2}][X_{K^*}^{\lambda}- z
Y_{K^*}^{\lambda}] ~.
\label{abxyvec}
\end{eqnarray}

There are several points to be deduced from the above results. First,
since $Y_{K^*}^{\lambda=0} \approx 0$, the coefficients $a$, $b$, $x$
and $y$ all have the same phase for a longitudinally-polarized
$K^*$. Thus all triple products involving a longitudinal $K^*$ are
expected to vanish. Furthermore, since $\varepsilon_V \cdot
p_{\Lambda_b} = 0$ for a transversely-polarized $K^*$, most of the
triple products in Eq.~(\ref{vecTPs}) are expected to vanish in the
SM. The only potential nonzero triple-product correlations are
\beq
2 \, {\rm Im} \left(x y^*\right) \epsilon_{\alpha\beta\mu\nu} \left[
\varepsilon_{K^*} \cdot s_p \, p_p^\alpha p_{\Lambda_b}^\beta
s_{\Lambda_b}^\mu \varepsilon_{K^*}^\nu + \varepsilon_{K^*} \cdot
s_{\Lambda_b} \, p_p^\alpha s_p^\beta \varepsilon_{K^*}^\mu
p_{\Lambda_b}^\nu \right] ~.
\eeq
Since these both require the measurement of all three spins, this
result is consistent with the results of Ref.~\cite{Wafia}. 

Second, and more importantly, both of these asymmetries only arise due
to the interference between the (small) dipole term
$Y_{K^*}^{\lambda=\perp}$ and the $V_{ub} V_{us}^*$ piece of
$X_{K^*}^{\lambda=\perp}$. Thus, by analogy with
Eq.~(\ref{Pestimate}), we estimate the size of the asymmetries to be
roughly
\beq
{2 {\rm Im} (xy^*) \over |x|^2 + |y|^2 + 2 {\rm Re}(xy^*)} \simeq {z
{\rm Im}(X_{K^*}^{\lambda=\perp} Y_{K^*}^{{\lambda=\perp}^*}) \over
|X_{K^*}^{\lambda=\perp}|^2} \simeq {z a_2 a_d \eta \lambda^2 \over
a_2^2 \lambda^4 + a_4^2} \sim 2\% ~,
\label{Vestimate}
\eeq
which would be very difficult to measure. (Essentially, the asymmetry
is reduced compared to that in $\Lambda_b \to p K^-$ by the factor $z
|a_d/a_6| = 0.11$.) Furthermore, the decay $\Lambda_b \to p K^{*-}$ is
dominated by the longitudinally-polarized $K^{*-}$; the rate for the
production of a transversely-polarized $K^{*-}$ is suppressed by the
factor $(m_{K^*}/E_{K^*})^2 = 0.1$.  Thus, even if the asymmetry were
larger, it would still be difficult to detect, given the small rate.

We therefore conclude that {\it any} measurement of a sizeable
triple-product asymmetry in the decay $\Lambda_b \to p K^{*-}$ is an
unequivocal signal of new physics \cite{newpaper}. (As noted in the
case of $\Lambda_b \to p K^{-}$ decay, if the measurement of the
proton polarization is difficult, one can consider a final state with
an excited nucleon such as $\Lambda_b \to N(1440) K^{*-}$. In this
case, the polarization of the $N(1440)$ can be determined from its
decay products. Alternatively, one can consider $\Xi_b \to \Sigma^+
K^{*-}$, for which the above conclusions should also hold.)

The decay $\Lambda_b \to p\rho^{-}$ is similar to $\Lambda_b \to p
K^{*-}$, and its amplitude can be obtained from Eqs.~(\ref{vector}),
(\ref{abxyvec}) and (\ref{pkstar}) with the replacements $V_{is} \to
V_{id}$ and $K^* \to \rho$. However, here too the asymmetry is
expected to be smaller than that in $\Lambda_b \to p \pi^-$ by the
factor $z |a_d/a_6| = 0.11$, yielding an asymmetry of less than 1\%.
Finally, the pure penguin process $\Lambda_b \to \Lambda \phi$, is
dominated by a single weak amplitude, so that all its triple-product
asymmetries vanish. (As mentioned earlier, the observation of the
decay $\Lambda_b \to \Sigma \phi$ would indicate the existence of
nonfactorizable contributions and the possible presence of a
significant $V_{ub}V_{us}^*$ piece in the amplitude.)

To summarize, we have examined the predictions of the standard model
for T-violating triple-product asymmetries in $\Lambda_b \to F_1 F_2$
decays, where $F_1$ is a light spin-$1\over 2$ baryon, and $F_2$ is a
pseudoscalar ($P$) or vector ($V$) meson. In $\Lambda_b \to F_1 P$
decays, there is only a triple product possible. In the rest frame of
the $\Lambda_b$, it takes the form ${\vec p}_{F_1} \cdot ({\vec
s}_{F_1} \times {\vec s}_{\Lambda_b})$, where ${\vec p}_{F_1}$ is the
3-momentum of the $F_1$, and ${\vec s}_{F_1}$ and ${\vec
s}_{\Lambda_b})$ are the spins of the $F_1$ and $\Lambda_b$,
respectively. On the other hand, in $\Lambda_b \to F_1 V$ decays,
since all three particles have a non-zero spin, there are several
possible triple products.

Using factorization, we find the following results. First, for
$\Lambda_b \to F_1 P$ decays, the SM predicts a large asymmetry ($\sim
18\%$) only for $\Lambda_b \to p K^-$. This is due to the presence of
the chiral enhancement term $m_K^2/(m_s+m_u)$ in the amplitude, which
compensates the $1/m_b$ suppression. The asymmetry in $\Lambda_b \to p
\pi^-$ is smaller ($\sim 5\%$), and for the decays $\Lambda_b \to
\Lambda \eta, \Lambda \eta'$, it is less than 1\%. Second, for
$\Lambda_b \to F_1 V$ decays with a transversely-polarized $V$, the
asymmetries are quite small: for $\Lambda_b \to p K^{*-}$ and
$\Lambda_b \to p \rho^-$ they are $O(1\%)$ and $<1\%$,
respectively. The asymmetries involving a longitudinally-polarized $V$
are expected to be roughly 15\% smaller than those for a
transversely-polarized $V$, so that they are effectively unmeasurable.
There are no asymmetries in $\Lambda_b \to \Lambda \phi$ since this
decay is dominated by a single weak amplitude. The fact that, within
the SM, the triple-product asymmetries in many decays are tiny
suggests that this is a good area to search for new physics.

\bigskip
\noindent
{\bf Acknowledgements}:
We thank S. Pakvasa and R. MacKenzie for helpful discussions. This
work was financially supported by NSERC of Canada.


\end{document}